\documentstyle[12pt]{article}
\begin{document}\thispagestyle{empty}\begin{flushright}
OUT--4102--67\\
hep-th/9612013\\
30 November 1996
            \end{flushright} \vspace*{2mm} \begin{center} {\Large\bf
Compact analytical form for non-zeta terms\\[5pt]
in critical exponents at order $1/N^3$}\vglue 10mm{\large{\bf
D.~J.~Broadhurst$            ^{1)}$}\vglue 4mm
Physics Department, Open University\\[3pt]
Milton Keynes MK7 6AA, UK           \vglue 6mm{\bf
A.~V.~Kotikov$               ^{2)}$}\vglue 4mm
Particle Physics Laboratory\\[3pt]
Joint Institute for Nuclear Research\\[3pt]
141980 Dubna, Russia}\end{center}\vfill\noindent{\bf Abstract}\quad
We simplify, to a single integral of dilogarithms, the least tractable
${\rm O}(1/N^3)$ contribution to the large-$N$ critical exponent $\eta$ of
the non-linear $\sigma$-model, and hence $\phi^4$-theory,
for any spacetime dimensionality, $D$. It is the sole generator of
irreducible multiple zeta values in $\varepsilon$-expansions with
$D=2-2\varepsilon$, for the $\sigma$-model, and $D=4-2\varepsilon$,
for $\phi^4$-theory. In both cases we confirm results of Broadhurst,
Gracey and Kreimer (BGK) that relate knots to counterterms.
The new compact form is much simpler than that of BGK.
It enables us to develop 8 new terms in the $\varepsilon$-expansion with
$D=3-2\varepsilon$. These involve {\em alternating\/}
Euler sums, for which the basis of irreducibles is larger.
We conclude that massless Feynman diagrams in odd
spacetime dimensions share the greater transcendental complexity of massive
diagrams in even dimensions, such as those contributing to the
electron's magnetic moment and the electroweak $\rho$-parameter.
Consequences for the perturbative sector of Chern-Simons theory
are discussed.                               \vfill\footnoterule\noindent
$^1$) email: D.Broadhurst@open.ac.uk\\
$^2$) email: kotikov@sunse.jinr.dubna.su
\newpage\setcounter{page}{1}
\newcommand{\df}[2]{\mbox{$\frac{#1}{#2}$}}
\newcommand{\Eq}[1]{~(\ref{#1})}
\newcommand{\Eqq}[2]{~(\ref{#1},\ref{#2})}
\newcommand{\ep}{\varepsilon}
\newcommand{\la}{\lambda}
\newcommand{\De}{\Delta}
\newcommand{\rd}{{\rm d}}
\newcommand{\hyp}{{}_3F_2}
\newcommand{\ps}[1]{\psi_{#1}}
\newcommand{\psp}[1]{\psi^\prime_{#1}}
\newcommand{\pspp}[1]{\psi^{\prime\prime}_{#1}}

\subsection*{1. Introduction}

Since the pioneering work of the St Petersburg group~\cite{VPH,Imu},
exploiting conformal invariance~\cite{AMP} of critical phenomena,
it was known that the ${\rm O}(1/N^3)$ term $\eta_3$
in the large-$N$ critical exponent $\eta$ of the non-linear $\sigma$-model,
or equivalently $\phi^4$-theory, in any number $D\equiv2\mu$ of
spacetime dimensions, derives its maximal complexity
from a single Feynman integral $I(\mu)$, defined below.
The situation is thoroughly reviewed
by Broadhurst, Gracey and Kreimer in~\cite{BGK} (hereafter BGK),
who first showed that the $\ep$-expansions
of $I(1-\ep)$ and $I(2-\ep)$, for the $\sigma$-model and $\phi^4$-theory,
entail double Euler sums~\cite{LE}, of the type
$\zeta(s,t)=\sum_{m>n>0}m^{-s}n^{-t}$.
The majority of such sums are conjectured~\cite{DZ1,DZ2,BBG,BBB,MEH,PETA,BBBR}
to be irreducible to the single sums $\zeta(r)=\sum_{n>0} n^{-r}$,
when $s$ and $t$ have the same parity, the weight $s+t$ exceeds 6,
and $s>t>1$.
Prior to BGK, the first double-sum irreducible, conjectured~\cite{1440} to be
at weight 8, had not been detected by large-$N$
studies~\cite{JAG,VDKS,KSV}.

Thus $I(\mu)$ provides important information~\cite{BGK} on the
mapping~\cite{DK1,DK2} between
positive knots~\cite{BK15,DK} and transcendental numbers~\cite{BBG,BG,EUL},
realized by the counterterms~\cite{BKP,BDK,DKT,DK3,DEM} of
perturbative quantum field theory,
in even numbers of spacetime dimensions, and most pertinently in the four
dimensions where particle physics is studied experimentally.
This connection arises from the
skeining~\cite{VJ} of link diagrams~\cite{DK1} that encode momentum
flow in Feynman diagrams. The field-theoretic connection
between $\zeta(2n+1)$ and the 2-braid torus knot~\cite{VJ} $(2n+1,2)$
is now well understood~\cite{DK1,DK2,DK3}.
Thanks to~\cite{Imu}, BGK were able to explore the field-theoretic
connection between 3-braid knots and irreducible double Euler sums,
at loop orders much higher than the 7 loops achieved in~\cite{BKP},
and hence for knots with many crossings and Euler sums of large weights.

The purpose of the present paper is to give a representation
of $I(\mu)$ that is considerably simpler than that achieved by BGK,
and to exploit it by addressing a further question in
knot/number/field theory~\cite{DK}:
what is the character of the transcendentals that emerge from massless,
single-scale Feynman diagrams in {\em odd} numbers of dimensions?

These possibilities arise from the work of one us (AVK) in~\cite{AVK},
following communication from the other (DJB) that BGK had succeeded in
reducing a large class of Feynman integrals, including $I(\mu)$, to
$\hyp$ series, of the type first revealed in~\cite{SJH}
and intensively studied in~\cite{OSC}. As a result,
an alternative route to $\hyp$ series was found in~\cite{AVK}, via
Gegenbauer-polynomial techniques~\cite{GPX}. While such techniques
are sometimes less efficient than the use of recurrence
relations~\cite{BGK,OSC}, and require considerable ingenuity to
convert~\cite{AVK} Gegenbauer double series to $\hyp$ series, they have
facilitated the current
project of studying the $\ep$-expansion of $I(\frac32-\ep)$, for which only
the leading~\cite{Imu} term, $I(\frac32)=-7\zeta(3)/2\zeta(2)+2\ln2$, was
previously known.
We shall show that higher terms in this $\ep$-expansion, involving
more loops in the regularization of three-dimensional Feynman
diagrams, entail irreducible {\em alternating\/} Euler sums.
In contrast, non-alternating sums, which we call multiple zeta
values (MZVs)~\cite{DZ2,PETA,BK15}, emerge in even dimensions.
Massless multiloop diagrams in odd dimensions thus have greater
analytical complexity than those in even dimensions. Indeed,
single-scale massless diagrams in three dimensions appear
to have the same character
as single-scale massive~\cite{EUL,OSC} diagrams in four dimensions.

In Section~2 we recall the definition~\cite{Imu} of $I(\mu)$ and give
the new result for it.
A derivation is sketched in Section~3. Sections~4 and~5 show how MZVs
and alternating sums emerge in even and odd dimensions, respectively.
Section~6 gives our conclusions.

\subsection*{2. New compact result for $I(\mu)$}

The parameter $\mu\equiv D/2$, used in~\cite{VPH,Imu},
is less convenient than $\la\equiv\mu-1$, which
arises as the exponent in the bare propagator,
$\left[1/(x-y)^2\right]^\la\equiv1/(x-y)^{2\la}$, of a scalar particle in
$D$-dimensional configuration space. Our aim is to expand
$I(\mu)=I(\lambda+1)$ around $\la=0$, and $\la=\frac12$,
corresponding to $D=2$ and $D=3$, respectively.
Expansions around larger integer values of $D$ may then be achieved by a
recurrence relation~\cite{BGK,KSV} that increases the
dimensionality by two units.

The definition of $I$ is~\cite{Imu}
\begin{equation}
I(\lambda+1)=\left.\frac{\rd}{\rd\De}\ln\Pi(\la,\De)\right|_{\De=0},
\label{def}
\end{equation}
where
\begin{equation}
\Pi(\la,\De)=\frac{{x^{2(\la+\De)}}}{\pi^D}\int\int\frac{\rd^D y \rd^D z}
{y^2z^2(x-y)^{2\la}(x-z)^{2\la}(y-z)^{2(\la+\De)}}\label{Pi}
\end{equation}
is a two-loop two-point integral, with three dressed propagators,
made dimensionless by the appropriate power of $x^2$.

The new result, obtained by AVK, is
\begin{equation}
I(\la+1)=\Psi(1)-\Psi(1-\la)+\frac{\Phi(\la)-\df13\Psi^{\prime\prime}(\la)
-\df{7}{24}\Psi^{\prime\prime}(1)}{\Psi^\prime(1)-\Psi^\prime(\la)}\,,
\label{ans}
\end{equation}
where
\begin{equation}
\Phi(\la)=4\int^1_0 d x \frac{x^{2\la-1}}{1-x^2}
\left\{{\rm Li}_2(-x)-{\rm Li}_2(-1)\right\}\,,\label{Phi}
\end{equation}
with $\Psi(x)=\Gamma^\prime(x)/\Gamma(x)$ and
${\rm Li}_2(x)=\sum_{n>0}x^n/ n^2$.

\subsection*{3. Method of derivation}

We begin by considering a more convenient Feynman integral, namely
\begin{equation}
J (\la,\De)=\frac{x^{2(2\la+\De-1)}}{\pi^D}\int\int\frac{\rd^D y \rd^D z}
{y^{2\la}z^{2\la}(x-y)^{2\la}(x-z)^{2\la}(y-z)^{2(1+\De)}}\,,\label{J}
\end{equation}
with a dressing only on the internal propagator.
It was studied in~\cite{AVK}, with the result
\begin{equation}
J(\la,\De)=\frac{2}{\la+\De-1}\,\frac{\Gamma(2\la+\De-1)}
{\Gamma^2(\la)\Gamma(2\la)}\,\frac{A(\la,\De)-B(\la,\De)}{\De},\label{Jis}
\end{equation}
where
\begin{eqnarray}
A(\la,\De)&=&\sum_{n=0}^{\infty}\frac{1}{n+\la+\De}\,
\frac{\Gamma (n+2\la)}{\Gamma (n+2\la+\De)}\,,\label{Ais}\\
B(\la,\De)&=&\frac{\Gamma(\la-\De)}{\Gamma(\la)}\frac{\pi}{\tan(\pi\De)}\,.
\label{Bis}
\end{eqnarray}
It is convenient to split $A=A_1+A_2$ into an easy and a difficult part:
\begin{eqnarray}
A_1(\la,\De)&=&\sum_{n=0}^{\infty}
\frac{1}{n+\la+\De}\,\frac{\Gamma(n+\la)}{\Gamma(n+\la+\De)}=
\frac{1}{\De}\,\frac{\Gamma(\la)}{\Gamma(\la+\De)}\,,\label{A1}\\
A_2(\la,\De)&=&\sum_{n=0}^{\infty}
\frac{1}{n+\la+\De}\left\{\frac{\Gamma(n+2\la)}{\Gamma(n+2\la+\De)}
-\frac{\Gamma(n+\la)}{\Gamma (n+\la+\De)}\right\}\,.\label{A2}
\end{eqnarray}
Only the first two terms of $A_2(\la,\De)=A_{2,1}(\la)\De
+\frac12A_{2,2}(\la)\De^2+{\rm O}(\De^3)$
are needed. By laborious procedures, whose detail would be inappropriate
here, these Taylor coefficients were obtained as
\begin{eqnarray}
A_{2,1}(\la)&=&\sum_{n=0}^{\infty}
\frac{\Psi(n+\la)-\Psi(n+2\la)}{n+\la}
=-\frac{\Psi^\prime(1)+\Psi^\prime(\la)}{2}\,,\label{A2p}\\
A_{2,2}(\la)&=&
3\Phi(\lambda)-\Psi^{\prime\prime}(\la)-\df{7}{8}\Psi^{\prime\prime}(1)
+\Psi(\la)\left\{\Psi^\prime(1)-\Psi^\prime(\la)\right\}\,,\label{A2pp}
\end{eqnarray}
where $\Phi$ is defined in\Eq{Phi}, and is the sole term that is not
reducible to polygammas.
Hence we are able to expand\Eq{J} to ${\rm O}(\De)$.

The transformation labelled ``$\to$'' in~\cite{VPH} then yields
\begin{equation}
\Pi(\la,\De)=\frac{\Gamma^2(\la)\Gamma(\la+\De)\Gamma(2-\De-\la)}
{\Gamma(1-\De)\Gamma(2\la+\De-1)}\,J(\la,\De)\,,\label{to}
\end{equation}
whose ${\rm O}(\De)$ term is needed in\Eq{def}.
Collecting terms, we obtain the advertised result\Eq{ans}.

\subsection*{4. MZVs in even dimensions}

It was found in~\cite{BGK,KSV} that $I(\la+1)$ is related to $I(\la)$
by a rather complex recurrence relation. The origin of this is immediately
apparent from\Eq{Phi}, which exposes the integral
at the heart of the intractability of $I(\la+1)$.
By expanding ${\rm Li}_2(-x)=\sum_{k>0} (-x)^k/k^2$, one obtains
the recurrence relation
\begin{equation}
\Phi(\la)-\Phi(\la+1)=4\sum_{k>0}\frac{(-1)^k}{k^2}
\left\{\frac{1}{2\la+k}-\frac{1}{2\la}\right\}
=\frac{\Psi(2\la+1)-\Psi(\la+1)}{\la^2}\,,
\label{rec}
\end{equation}
which shifts $D$ by 2.
Its solution is
\begin{equation}
\Phi(\la)=4\sum_{k>0}\frac{(-1)^k}{k^2}\sum_{n\ge0}
\left\{\frac{1}{2\la+k+n}-\frac{1}{2\la+n}\right\}\frac{1+(-1)^n}{2}\,,
\label{sol}
\end{equation}
which may be obtained directly from\Eq{Phi} by
expanding both $1/(1-x^2)$ and ${\rm Li}_2(-x)$.

It is apparent from\Eq{sol} that $\ep$-expansions
in $D=2-2\ep$ and $D=3-2\ep$ dimensions may be obtained in terms of
alternating Euler sums. The results are
\begin{eqnarray}
\Phi(-\ep)+\frac{\zeta(2)}{\ep}&=&
-3\zeta(2)\left\{\ln2+\Psi(1-\ep)-\Psi(1-2\ep)\right\}
-2\sum_{r>0}(2\ep)^{r-1}T_{+}(2,r)\,,
\label{d2}\\
\Phi(\df12-\ep)&=&
\phantom{-}3\zeta(2)\left\{\ln2+\Psi(1-\ep)-\Psi(1-2\ep)\right\}
-2\sum_{r>0}(2\ep)^{r-1}T_{-}(2,r)\,,
\label{d3}
\end{eqnarray}
where
\begin{equation}
T_{\pm}(s,t)=\sum_{m>n>0}\frac{(-1)^m\pm(-1)^n}{m^s n^t}\,.
\label{Tpm}
\end{equation}
For $D=4-2\ep$ one has merely to set $\la=-\ep$ in\Eq{rec}
and use\Eq{d2}.

The simplicity of\Eqq{d2}{d3} relies on the use of
{\em alternating\/} sums in\Eq{Tpm}. However, $\Phi(-\ep)$ should be
expandable in terms of non-alternating sums (i.e.\ MZVs) since that was the
finding of BGK, where
a result was obtained for $I(1-\ep)$ in terms of the double sum
\begin{equation}
S_+(\ep)\equiv\sum_{m>n>0}\frac{\ep^3}{(m+\ep)^2(n-\ep)}
\,+\,(\ep\to-\ep)\,{}=\sum_{s,t>0}(s-1)\zeta(s,t)
\left\{(-1)^s\pm(-1)^t\right\}
\ep^{s+t}
\label{Sp}
\end{equation}
and the polygammas
$\psi_p^{(n)}\equiv\left\{\partial/\partial p\right\}^{n+1}
\ln\Gamma(1+p\,\ep)=\ep^{n+1}\Psi^{(n)}(1+p\,\ep)$.

Recasting the result of BGK in terms of\Eq{ans}, we find that
\begin{eqnarray}
\ep^3\Phi(-\ep)&=&\df12S_+(\ep)
+\df18(3\ps1-7\ps{-1}-2\ps2+6\ps{-2}+6\psp0-3\psp1-5\psp{-1})
\nonumber\\&&{}
-\df1{16}(\pspp1-3\pspp{-1})
-\df32\psp0   (\ps1- \ps{-1}-\ps2+ \ps{-2})
\nonumber\\&&{}
+\df14\psp1   (\ps1+ \ps{-1}-\ps2- \ps{-2})
+\df14\psp{-1}(\ps1-3\ps{-1}-\ps2+3\ps{-2})\,,
\label{dict}
\end{eqnarray}
which we have verified, up to terms of ${\rm O}(\ep^{19})$,
using a suitable basis~\cite{EUL} for all double Euler sums up to weight 19,
obtained by methods developed in~\cite{BBG} and augmented in~\cite{EUL}.
A convenient {\bf Q}-basis for double sums, which is
conjectured~\cite{EUL} to be minimal, is formed by
$\ln2$, $\pi^2$, $\{\zeta(2a+1)\mid a>0\}$
and the alternating double sums $\{U(2a+1,2b+1)\mid a>b\ge0\}$,
of the form~\cite{1440}
$U(s,t)=\sum_{m>n} (-1)^{m+n} m^{-s}n^{-t}$. All 3698 convergent double
Euler sums with weights up to 44 have been expressed in this basis.
Moreover high-precision lattice methods~\cite{EUL} reveal no rational relations
between the basis elements. It was thus a simple matter of programming
to demonstrate the agreement of\Eq{d2} with\Eq{dict},
up to terms of ${\rm O}(\ep^{19})$ in\Eq{dict}.

The circumstance that\Eq{d2} yields only MZVs
may be restated, using the identity
\begin{equation}
\zeta(s,t)+U(s,t)+T_{+}(s,t)=2^{2-s-t}\zeta(s,t)\,,
\label{triv}
\end{equation}
which is obtained by retaining only even values of $m$ and $n$
in $\sum_{m>n>0}m^{-s}n^{-t}$. {}From\Eq{triv} it follows that
$T_+(2,r)$, in\Eq{d2}, is expressible in terms of MZVs if and only if $U(2,r)$
is so expressible. AVK has devised an elementary proof of the latter
proposition.

Thus it may be seen that the simplicity of\Eqq{ans}{d2} is won at some price:
it disguises the MZV-content of the even-dimensional case, discovered by
BGK. However, there was a consequent unexpected gift from field theory to
number theory. {}From the equivalence of the present result for $I(\mu)$ with
that of BGK, one of us (DJB) inferred a compact single formula
(subsequently proven, from first principles, by Roland Girgensohn) that gives
{\em all\/} the reductions of odd-weight double sums to single sums
and their products. The four possible choices of sign were not treated on an
equal footing in~\cite{BBG}, where the reduction of alternating sums was left
in a `somewhat more implicit state', compared with the non-alternating
case. The reader is referred to a recent
compendium~\cite{BBB} of results for Euler sums, where this field-theory
byproduct appears as Equation~(74), in a notation which unifies all
8 cases that result from the four possible choices of sign, and the
two possible choices of the opposite parities of the exponents
in odd-weight double sums.

\subsection*{5. Irreducible alternating sums in odd dimensions}

For expansion\Eq{d3}, in $D=3-2\ep$ dimensions, MZVs are insufficient.
The odd powers of $\ep$ in\Eq{d3} entail $T_{-}(2,2r)$, which
is a genuinely new double-sum transcendental.
However, study of all the cases with weights up to 44 reveals
that it may always be expressed in terms of MZVs and the single
alternating sum $U(2r+1,1)$, with a coefficient that follows a
regular pattern.
Specifically, we obtained the following $\ep$-expansion for $D=3-2\ep$:
\begin{equation}
I(\df32-\ep)=\frac{-7\zeta(3)+4\zeta(2)\ln2+\sum_{n>3}Y_n\ep^{n-3}}
{2\zeta(2)+\sum_{n>2}(n-1)(2^n-1)\zeta(n)\ep^{n-2}}\,,
\label{Ydef}
\end{equation}
where only the leading~\cite{Imu} term, $I(\frac32)=-7\zeta(3)/2\zeta(2)
+2\ln2$, was previously known. Developing the numerator to weight 11,
we obtain
\begin{eqnarray}
Y_4 &=&-\df{53}{2}\zeta(4)+16U(3,1)\nonumber\\
Y_5 &=&-42\zeta(3)\zeta(2)-\df{217}{2}\zeta(5)+90\zeta(4)\ln2\nonumber\\
Y_6 &=&-456\zeta(6)+29\zeta^2(3)+128U(5,1)\nonumber\\
Y_7 &=&7\zeta(3)\zeta(4)-434\zeta(5)\zeta(2)-\df{3937}{4}\zeta(7)
        +630\zeta(6)\ln2\nonumber\\
Y_8 &=&-153\zeta(6,2)-\df{18321}{4}\zeta(8)+822\zeta(5)\zeta(3)+768U(7,1)
       \nonumber\\
Y_9 &=&889\zeta(3)\zeta(6)-1333\zeta(5)\zeta(4)-2794\zeta(7)\zeta(2)
      -\df{20951}{3}\zeta(9)+3570\zeta(8)\ln2\nonumber\\
Y_{10}&=&-591\zeta(8,2)-\df{64265}{2}\zeta(10)+2176\zeta^2(5)
       +4340\zeta(7)\zeta(3)+4096U(9,1)\nonumber\\
Y_{11}&=&7147\zeta(3)\zeta(8)-1891\zeta(5)\zeta(6)-11049\zeta(7)\zeta(4)
       -15330\zeta(9)\zeta(2)
        \nonumber\\&&{}
        -\df{173995}{4}\zeta(11)+18414\zeta(10)\ln2
\label{Y}
\end{eqnarray}
with $U(2s-1,1)$ appearing in $Y_{2s}$, with coefficient
$(s-1)4^s$, and $\zeta(2s)\ln2$ in $Y_{2s+1}$, with coefficient
$2(2s-1)(4^s-1)$, for $s>1$. All other contributions are MZVs, though
some are irreducible to single sums, as in even dimensions.

We conclude that the appearance of $\ln2$ in the strictly three-dimensional
result of~\cite{Imu}, was merely the opening of the floodgates
to further non-MZV terms.
At weight 4, one encounters~\cite{EUL}:
\begin{equation}
U(3,1)\equiv\sum_{m>n>0}\frac{(-1)^{m+n}}{m^3n}
=\df12\zeta(4)-2\left\{{\rm Li}_4(\df12)
+\df{1}{24}\ln^22\left(\ln^22-\pi^2\right)\right\}\,,\label{U31}
\end{equation}
where the non-MZV polylogarithm ${\rm Li}_4(\frac12)$
occurs with precisely the same~\cite{EUL} combination of
$(\ln2)^4$ and $(\pi\ln2)^2$ as in the massive four-dimensional three-loop
results for the anomalous magnetic moment of the electron~\cite{LR} and
the $\rho$-parameter~\cite{AFMT} of electroweak theory.
Only the combinations of $U(3,1)$ with $\zeta(4)=\frac{\pi^4}{90}$
differ from that in $Y_4$ of\Eq{Y}. This is a remarkable circumstance,
giving concrete support to the idea, motivated by hypergeometric
analysis~\cite{EUL,OSC}, that massless diagrams in odd dimensions
lead to results that are transcendentally more complex
than those for massless
even-dimensional diagrams, and are closely akin to {\em massive\/}
even-dimensional results. Nor do simple polylogarithms
exhaust the novelty; at weight 6 there is no~\cite{BBG,EUL} known
integer relation between
$U(5,1)$, ${\rm Li}_6(\frac12)$ and simpler polylogs.

\subsection*{6. Conclusions}

In conclusion, we have obtained and exploited a remarkably simple
representation\Eqq{ans}{Phi} for the least tractable integral $I(\mu)$
in the large-$N$ critical exponent $\eta$~\cite{Imu} at order $1/N^3$.
The even-dimensional $\ep$-expansions of BGK~\cite{BGK} were confirmed
up to weight 19, making the probability of error in the new result\Eq{ans},
or in the BGK result, negligible. Checking their equivalence
had a beneficial spin-off for number theory~\cite{BBB}. It is remarkable how
closely the massless three-dimensional results mimic massive
four-dimensional results~\cite{LR,AFMT}. The origin of this is clear:
alternating Euler sums are the common new ingredient, and\Eq{U31}
is the only~\cite{EUL} possible irreducible depth-two interloper at weight 4,
if Euler sums exhaust the transcendentals from single-scale diagrams, at
this weight.

A large questions remains: are alternating Euler sums the numbers assigned
to knots via counterterms of odd-dimensional field theories? It must be
emphasized that we do not yet know the answer. There are manifold
successes~\cite{BGK,BK15,BKP,BDK,DKT,DK3,DEM} of using knots~\cite{DK1,DK2,DK} to
relate the transcendentality content of even-dimensional counterterms to
the skeining~\cite{VJ} of link diagrams that encode momentum flow. Very
recent work~\cite{4TR,BK4} suggests an underlying weight system associated
with four-term~\cite{BN} relations. Yet it is vital to remember that all
this was achieved by the study of renormalizable field theories in their
critical dimensions, which were {\em even\/} in the cases studied so far.
The all-order results of~\cite{Imu}, for critical exponents, have
predictive content for the perturbative sectors of the $\sigma$-model and
$\phi^4$-theory at their critical dimensions of $D=2$ and $D=4$,
respectively, where counterterms, from nullified diagrams, are the simplest
possible field-theoretic constructs. Until one studies a theory whose
critical dimension is $D=3$, one has no right to associate knots with
alternating Euler sums, via counterterms.

Surprisingly, there appears to be no literature on $\phi^6$-theory,
in its critical dimension $D=3$, beyond the two-loop~\cite{Tan} level.
Two-loop counterterms are trivial from the point view of
knot theory~\cite{DK1,DK2},
as the trefoil knot first occurs at three loops. Gratifyingly,
they also appear to be trivial from the point of view of number theory,
with a rational two-loop beta-function in~\cite{Tan},
agreeing with the expectations of~\cite{DK1,DK2}. In some
renormalization schemes, subdivergences~\cite{DK3}
may generate $\pi^2$-terms~\cite{Kaz} in anomalous dimensions, at $D=3$,
corresponding to framing dependence
in knot theory~\cite{DK}. There is a
four-loop~\cite{BB} analysis of diagrams that are logarithmically divergent
for $D=3$, and this indeed yields $\pi^2$ terms.
In even dimensions~\cite{DK3}, by contrast,
the first framing dependence shows up at the level of $\zeta(4)=\pi^4/90$,
in the scheme-dependence of anomalous dimensions.
We commend study of $\phi^6$-theory at the three-loop level, and preferably
beyond. It may provide the first non-trivial results on the mapping from
knots to transcendental numbers via the counterterms of a
field theory whose critical dimension is odd.

In the meantime, our $\ep$-expansion\Eqq{Ydef}{Y}, for
$D=3-2\ep$, fully confirms the hypergeometric expectation of~\cite{OSC}
that the transcendental complexity of massless Feynman integrals in three
dimensions is comparable to that of massive~\cite{LR,AFMT} diagrams in four
dimensions and hence entails alternating~\cite{EUL} Euler sums. This
is sobering news for colleagues seeking to push back the computational
frontier in the perturbative sector~\cite{Tan} of Chern-Simons theory.
At present
this lags far behind the 7-loop~\cite{BKP} level, achieved for
$\phi^4$-theory with $D=4$, in an analysis that spectacularly confirmed
Kreimer's predictions~\cite{DK1,DK2} for the fascinating nexus of
knot/number/field theory~\cite{DK}, whose study has advanced with great
rapidity~\cite{BGK,DZ2,BBB,MEH,PETA,BBBR,BK15,EUL,DK3,DEM,4TR,BK4}
in recent months.

\subsection*{Acknowledgements}
AVK is grateful to Prof.\ A.N.\ Vasil'ev, for stimulating the quest
for\Eq{ans}, and also to Prof.\ D.I.\ Kazakov, Dr.\ N.A. Kivel, and Dr.\
A.S.\ Stepanenko, for discussion. DJB thanks Jon Borwein, David Bradley,
Roland Girgensohn and Don Zagier, for discussions on number theory, Bob
Delbourgo and John Gracey, for discussions on field theory, and Dirk
Kreimer, for knotting the two together during a collaboration
on~\cite{BGK,BK15,BK4} at the University of Tasmania, generously hosted
by Bob Delbourgo in July and August.

\raggedright

\end{document}